\documentclass[pre,aps,floats,superscriptaddress,floatfix]{revtex4-2}
\usepackage{amssymb,amsmath}
\usepackage{amsmath,amssymb}
\usepackage{graphicx}
\usepackage{psfrag}
\usepackage{color}
\usepackage{dcolumn}
\usepackage{bm}
\usepackage[normalem]{ulem}
\usepackage{gensymb}
\usepackage{natbib}
\usepackage{mathtools}

\begin{document}

\title{Achieving Maximum Utilization in Optimal Time for Learning or Convergence in the Kolkata Paise Restaurant Problem}
\author{Aniruddha Biswas}
\email{aniruddha.biswas@ltimindtree.com}
 \affiliation{LTIMindtree Limited, Kolkata, India}
\author{Antika Sinha}
\email{antikasinha@gmail.com}
 \affiliation{Department of Computer Science, Asutosh College, Kolkata-700026, India}
\author{Bikas K. Chakrabarti}
 \email{bikask.chakrabarti@saha.ac.in}
\affiliation{Saha Institute of Nuclear Physics, Kolkata-700064, India}
\affiliation{Economic Research Unit, Indian Statistical Institute, Kolkata-700108, India}

\date{\today}

\begin{abstract}
In the original version of the Kolkata Paise
Restaurant (KPR) problem, where each of the  $N$
agents (or players) choose independently every
day (updating their strategy based on past
experience of failures) among the $N$ restaurants,
where he/she will be alone or lucky enough to be
picked up  randomly from the crowd who arrived at
that  restaurant that day, to get the only food
plate served there. The objective of the agents
are to learn themselves in the minimum (learning)
time to have maximum success or utilization
probability ($f$). A dictator can easily solve
the problem with $f = 1$ in no time, by asking
every one to form a queue and go to the respective
restaurant, resulting in no fluctuation and full
utilization from the first day (convergence time
$\tau = 0$). It has already been shown that if
 each agent chooses randomly the restaurants, $f = 1 - e^{-1}
\simeq 0.63$ (where $e \simeq 2.718$  denotes the Euler
number) in zero time ($\tau = 0$). With the only available information about
yesterday's crowd size in the restaurant visited by the
agent (as assumed for the rest of the strategies studied 
here), the crowd avoiding (CA) strategies can give higher values of $f$ but also of $\tau$.   
Several numerical studies
of modified  learning strategies actually
indicated increased value of $f = 1 - \alpha$
for $\alpha \to 0$, with $\tau \sim 1/\alpha$.
We show here using Monte Carlo technique, a
modified Greedy Crowd Avoiding (GCA) Strategy can
 assure full 
 utilization ($f = 1$) in convergence time $\tau \simeq eN$,
with of course non-zero probability for an even larger convergence
time. All these observations
suggest that the strategies with
single step memory of the individuals can never
collectively achieve full utilization ($f = 1$)
in finite convergence time and perhaps the
maximum possible utilization that can be
achieved is about eighty percent ($f \simeq
0.80$) in an optimal time $\tau$ of order
ten, even when $N$ the number of customers or
of the restaurants goes to infinity.

\end{abstract}

\maketitle

\section{Introduction} 
Since early nineteen hundred, through
seventies, Kolkata had very cheap fixed price `Paise
Restaurants' (see e.g.,~\cite{kishan2021pice}), also called the `Paise
Hotels' (Paise is, rather was, the smallest
Indian coin). Because of the extremely low
price of their food and wide availability
throughout the city, such `Kolkata Paise
Restaurants' were very popular among the
daily  laborers coming everyday to the city
from the neighborhoods. During lunch hours,
these laborers used to  walk down (to save
the transport costs) from their place of work
to one of these restaurants. These restaurants
would prepare every day a (small) number of such
cheap dishes, sold at a fixed price (Paise). If
several groups of laborers would arrive any day
to the same restaurant, only one group would get
their lunch, and others would be refused. They would then miss their lunch that day. There
were no cheap communication means (like mobile
phones) for prior consultations, and  walking
down to the next restaurant would mean failing
to report back to work on time! To complicate
this large scale (many-body) dynamic collective
learning and decision-making problem, there were
indeed some well-known rankings of these
restaurants, as some of them would offer tastier
items compared to the others (at the same cost,
Paise, of course), and people would prefer to
choose the higher rank of the restaurant, if not
crowded! This `mismatch' of the choice and the
consequent decision not only creates inconvenience
for the prospective customers (going without lunch),
would also mean `social wastage' (excess unconsumed
food, services, or supplies somewhere). A similar
problem arises when the public administration plans
and provides, say, hospitals (beds) in different
localities, but the local patients prefer `better'
perceived hospitals elsewhere. These ‘outsider’
patients would then have to choose other suitable
hospitals elsewhere. Unavailability of the hospital
beds in the over-crowded hospitals may be considered
as insufficient service provided by the
administration, and consequently the unattended
potential services  will be considered as social
wastage.

In such many-player games, where each one tries to
learn independently from the past experience (and
consequent changes in their strategies), the
objective of each player is to maximize the his/her
chance to available resources in minimum time of
learning independently through the past failures
and consequent changes in the strategies of each.
Statistically, best  solution for individuals here
is also best for the society (less over all wastage
of resources achieved at the earliest). Note that
a dictated solution (provided by a non-playing
dictator) trivially achieves 100\% utilization of
resources (food for every one), and that too in no
time (from the very first day), by asking the
players to form a queue and follow their number
in the queue to go to the corresponding restaurant.
Even in cases of commonly agreed ranks of the restaurants, the dictated solution (asking to form a periodic chain or each person
   is asked to go to uniquely assigned restaurant, and
   then every one move to the next rank each day)
   achieves, again trivially, the
most efficient solution to the problem. The problem
comes when each player or customer decides for his
or her own.

The Kolkata Paise Restaurant (KPR) problem is a
repeated game, played among a large number ($N$)
of players or agents having no simultaneous
communication or interaction among themselves. 
The prospective players (customers/agents)
choose from the restaurants each day (discrete
time $t$) in parallel, based
on the past (crowd) information and their
own (evolved or learned) strategies. There is
no budget constraint to restrict the choice.
For simplicity, we may assume that each restaurant
can serve only one customer (generalization to any
fixed number of daily services for each would not
change the complexion of the problem or game) and
the total number of prospective customers on each
day is the same as the number $N$ of restaurants.
So, every one in principle can get a meal every
day. If, however, more than one customer arrives
at any restaurant on the same day, one of them is
randomly chosen and is served, and the rest arriving at 
that restaurant do not get a meal that day. Information
regarding the crowd sizes for the earlier days (up
to a finite memory size) is made available to
everyone. Each day, based on own learning and the
self-learned or evolved strategies, each customer
chooses a restaurant independent of the others. One
analyzes the steady state (large $t$) distribution
(in the large $N$ limit) of the fraction ($f$,
called the utilization fraction) of the agents
getting their meal or restaurants getting at least
one customer on average everyday, and studies how
to increase $f$ by collective learning by the
agents from their own previous experience (see
e.g.,~\cite{ghosh2011wolfram} ).

The KPR problem or game was first introduced in~\cite{chakrabarti2009kolkata}, by extending the many-agent Minority Game~\cite{challet2004minority} 
where each agent has two choices. As mentioned already, in KPR each of the $N$
agents have macroscopic number of choices ($M$) in
general, though in most of the studies of the
current literature, including this study, each
of the $N$ agents has got $N$ ($M=N$) choices so that
every one, in principle, can be  offered the
service (see e.g.,~\cite{chakraborti2015statistical,chakrabarti2021development,chakrabarti2022stochastic,ghosh2010statistics,chakrabarti2017econophysics} for reviews).

Following the initial formulation~\cite{chakrabarti2009kolkata}, the search
for strategies to increase the utilization
fraction ($f$) using the past crowd
information~\cite{ghosh2010statistics,ghosh2017emergence,sinha2020phase} (even in the case where
the number of agents and restaurants are not
identical~\cite{sinha2020phase}, though both are macroscopic),
its econophysical aspects~\cite{tamir2018econophysics}, analysis of
some of its game theoretic aspects~\cite{banerjee2018theeconomics,sharma2018thesaga},
utility in job scheduling in the context
of  Internet of Things~\cite{park2017kolkata} and for efficient
computation~\cite{picano2023efficient} employing the KPR algorithms
 had been studied and reported. 
 Beginning with some important early attempts to generalize
for quantum games, the quantum KPR games had been proposed~\cite{sharif2012strategies,ramzan2013three,sharif2013introduction}, which could be solved so far for three players only
(see~\cite{chakrabarti2017econophysics} for a very detailed review, and~\cite{chakrabarti2022stochastic} for a recent
review of these studies on quantum KPR games).

Some extensions for practical applications of the KPR game
has been to accommodate optimization not only by the agents
but also by the restaurants as well, to study the cases
of vehicles for hire problem~\cite{martin2019,martin2017vehicle,yang2018mean}. One of them~\cite{martin2017vehicle} discusses the inefficiencies of taxi drivers choosing their
own locations. This extended KPR model (where the taxis
which are equivalent to the restaurants here, can also
optimize like the prospective passengers or agents) uses
similar game theory to analyze how drivers strategically
choose the locations. The goal is to maximize driver
utilization and passenger waiting times.~\cite{yang2018mean} proposes
a mean field theory to analyze such optimization problems.
These studies propose several strategies for drivers to
choose locations, and finds that a mix of random and
strategic choices can lead to the highest utilization.

Also to map the KPR problem to $N$ city $N$ traveling
salesmen problem, one can allow each agent to make a few
additional cost-dependent further decisions or search in
the local neighborhood of the crowded choice~\cite{kastampolidou2022distributed} already
made. It assumes restaurants to be uniformly distributed
throughout the city and allows the agents to visit multiple
restaurants within their lunch break. This gives a new
perspective on the KPR problem, based on the Traveling
Salesman Problem algorithms to guide the agents in choosing
restaurants. This strategy prioritizes maximizing restaurant
utilization while still allowing agents to reach restaurants
within the time limit. It also allows the possibility for
agents to visit multiple restaurants if their initial
choice is unavailable. This increases the overall success
rate and reduces frustrations compared to the traditional
KPR assumptions.

In another recent version of KPR by allowing formation
of ``dining clubs''~\cite{harlalka2023stability} by availing the advantage of
their prior booking service at additional cost of
membership (fees, etc), in the KPR games, one can
achieve, under certain conditions, stability even when
players are allowed to cheat. The paper also studies the
effects of cheating on the size of dining clubs. The paper
shows that cheating can lead to smaller dining clubs. The
findings of the paper have implications for the design of
online marketplaces and other systems where cooperation
is important. This leads to intriguing and potentially
important applications of KPR algorithms to computer
science and management problems.

For the original version~\cite{ghosh2011wolfram,chakrabarti2009kolkata,ghosh2010statistics,sinha2020phase} of the
KPR problem, where $N$ agents each decide
independently (updating the strategy based
on past failures) every day choose the
restaurant, and if more than one chooses
the same restaurant, one is picked up
randomly from among the crowd and others
missing the meal that day, there had been~\cite{ghosh2010statistics} a few interesting  results (mostly numerically). We intend to a present here, again numerically, some confirmatory results for this
game also.

As mentioned already, a dictated solution
of the KPR problem, where each agent is
asked to form a (moving) queue and go to
the restaurant corresponding to the order
in the queue, helping every one to get food
(resource utilization fraction $f = 1$) from
day one ($t = 1$, convergence time $\tau
= 0$). The problem becomes nontrivial when
each agent tries to decide on his/her own and
resources (food dishes here) are limited;
number of dishes here are the same ($N$) as
the number of agents. Also the learning or
convergence time $\tau$ for reaching  the
best possible solution must be finite
(independent of $N$).

As shown already~\cite{chakrabarti2009kolkata}, for random choice of
the restaurant by any agent gives the
utilization fraction $f = 1 - exp(-1)
\simeq 0.63$ and, as there is no role of
memory, the convergence time $\tau = 0$.
With one step memory, each agent remembers,
say, the crowd size $n_k (t-1)$ at the
$k$th restaurant chosen and arrived at
 yesterday (with $\sum_{k=1}^{N}$, $n_{k(t)}
 = N$). If the agent chooses the same $k$th
restaurant today with probability $P_k(t)$
inversely dependent on the crowd size
$n_k(t-1)$ there yesterday ($P_k(t) =
1/[n_k(t-1)]^{\alpha}$, with $\alpha > 0$)
and chooses any other of the $N-1$
restaurants with probability $[(1- P_k(t))
/(N-1)]$, then the previous Monte Carlo
studies~\cite{ghosh2010statistics,sinha2020phase} suggested $f \simeq 0.80$
and average $\tau \le 10$ (independent of
$N$ for $\alpha = 1$; all studies with $N <
10^5$). For $\alpha$ approaching zero, the
Mote Carlo study in~\cite{sinha2020phase} indicated that while  the utilization fraction $f$
approaches unity as $1-\alpha$, while the convergence time
tends to diverge as $1/\alpha$ (independent of $N$).
Here in this paper, we will show, mostly
numerically, another efficient utilization
($f = 1$) algorithm, for which the
convergence time $\tau$ however grows linearly
with the system size $N$. 
This perhaps indicates the incompatibility of efficient
learning ($f \rightarrow 1$) in finite convergence time 
$\tau$ for infinite system size $N$ in
the Kolkata Paise Restaurant problem, where
each player learns and chooses independently.
As is seen, the dictatorial solution easily
achieves the compatibility, even for more
involved version of the KPR problem.

\begin{figure}[!htbp]
\includegraphics[width=12.5cm]{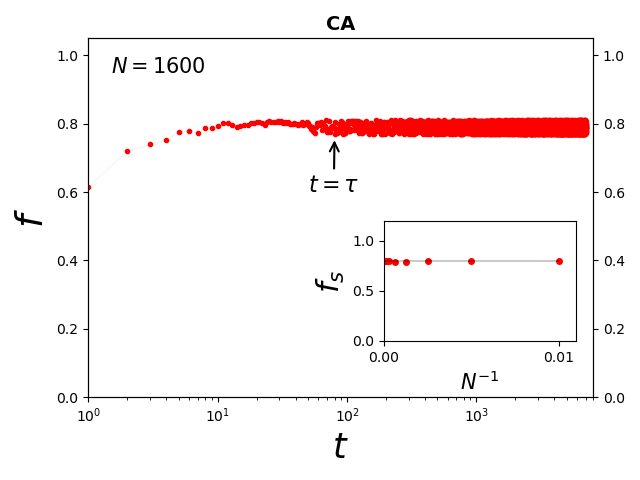}
 \caption{ Fraction $f$ of people getting food
(same as the utilization fraction $f$ of the
restaurants, getting a customer) as function
of (learning) time or day $t$, for a typical
Monte Carlo run of CA case with $N = 1600$,
the number of restaurants or of prospective
customers. The convergence time $\tau$ is also  
indicated, where $f$ saturates to an average value
$f_s$ and fluctuates around that value. The inset
shows the dependence of the
saturation value $f_s \simeq 0.80$ of the
utilization fraction $f$ as function of $N$
(up to $N = 51200$) and extrapolated to
$N \to \infty$.  }
 \label{fig-fasim}
\end{figure}

\section{Proposed learning algorithm and Monte
Carlo results}

We consider here a modification of the
learning version, studied earlier in~\cite{ghosh2010statistics},
where each agent or player goes back to the
same restaurant $k$ as chosen yesterday with
probability $P_k(t) = 1/[n_k(t-1)]$ and the
rest of the $N-1$ restaurants with equal
probability [$(1-P_k(t-1))/(N-1)$], where
$n_k(t)$ denotes the crowd size at the
$k$th restaurant on day $t$. Here we just
modify the algorithm such that out of the
$n_k(t-1)$ agents who reached the $k$th
restaurant yesterday, the one who was
picked up by the restaurant yesterday goes
with certainty to the same $k$th restaurant
next day, while others of the $n_k(t-1)$
agents go back to probability $P_k(t)$ and
to any of the other $N-1$ restaurants with
probability [$(1-P_k(t-1))/(N-1)$]. 
Here we study, using the Monte Carlo method, the
utilization fraction $f$ and the convergence time
$\tau$, for $N$ restaurants and $N$ agents with
$N$ = 100, 200, 400, 800, 1600, 3200, 6400, 12800, 25600 and 51200 
 where each agent is following the above mentioned crowd avoiding strategies.
The typical averaging is done over 30 runs for
each value of $N$.

\begin{figure}[!htbp]
\includegraphics[width=12.5cm]{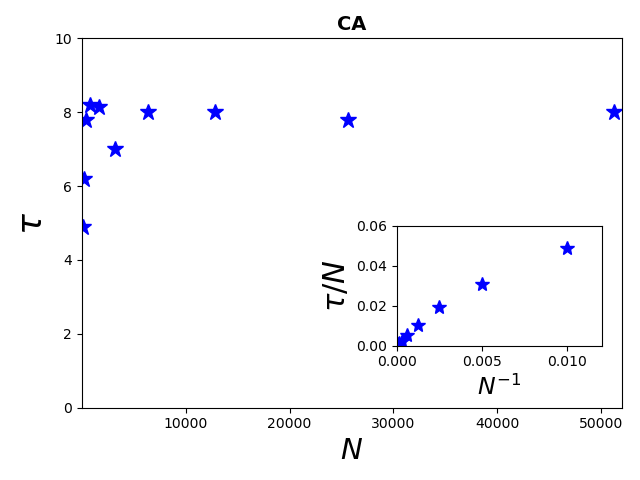}
 \caption{ 
 Convergence time $\tau \simeq 8$ as
function of $N$ in the CA case for $N =$ 100,
200, 400, 800, ...  up to $N = 51200$. Inset shows
that $\tau$ seems to  remain finite (less
than 10) even when extrapolated up to
$N \to \infty$.  }
 \label{fig-tasim}
\end{figure}

In GCA, this assures eventual full utilization
$f = 1$. Since once any restaurant gets occupied, it
remains occupied for all the successive (later) days,
though not necessarily by the same agent. However, this
limit of full utilization can only be achieved in $N$
order time; more precisely $\tau  = eN$, where
$e \simeq $ 2.71828 denotes the Euler number. This is
because, for random choice, the appropriate
probability~\cite{chakrabarti2009kolkata} $P(n)$ of choosing one restaurant by  
$n (\le N)$ players is given
by $P(n)= (1/n!)\exp(-1)$, for $N \to \infty$.
For $n = 1$ at any of the restaurants, the
value of $P(1)$ will be $e^{-1}$, giving the
required time $\tau/N = e$ for any of the
restaurants.

\begin{figure}[!htbp]
\includegraphics[width=12.5cm]{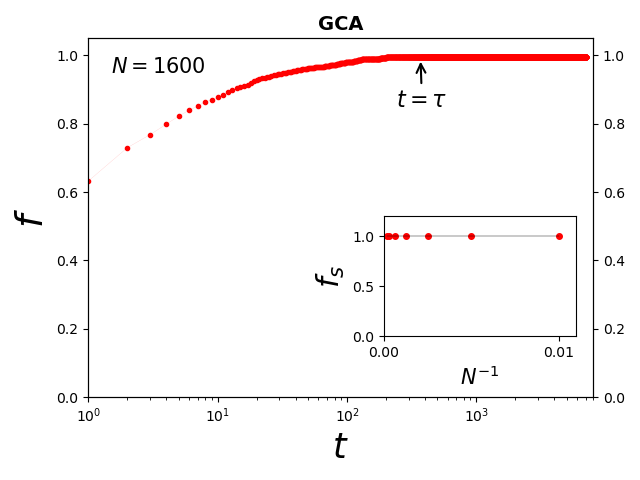}
 \caption{ 
 Utilization fraction $f$  as function
of (learning) time or day $t$, for a typical
Monte Carlo run of the  GCA case with
$N = 1600$. The convergence time $\tau$ is
also indicated, where $f$ value saturates
to unity. The inset shows that this full
utilization ($f_s =1$) occurs for all values
of $N$ (data here are up to $N = 6400$).  }
 \label{fig-fanir}
\end{figure}

\begin{figure}[!htbp]
\includegraphics[width=12.5cm]{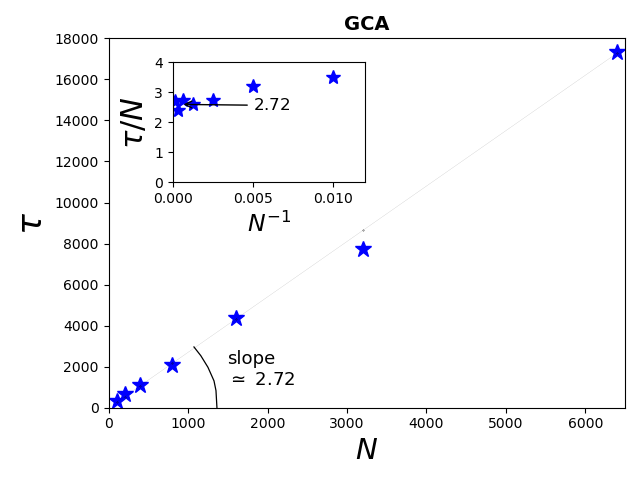}
 \caption{ 
 Convergence
time $\tau$ as function of $N$ in the GCA
case for $N$ up to 6400. Inset shows
that $\tau = eN$, where $e \simeq 2.718$
denotes the Euler number, seems to fit well
in the $N \to \infty$ limit.  }
 \label{fig-tanir}
\end{figure}

Let us now consider the results for two specific
cases or strategies defined earlier: Crowd Avoiding
(CA)  and Greedy Crowd Avoiding (GCA).

    a) For CA, the agent chooses the same $k$th
restaurant today with probability $P_k(t)$ inversely
proportional to the crowd size $n_k(t-1)$ there
yesterday: $P_k(t) = 1/[n_k(t-1)]$, and
chooses any other of the $N-1$ restaurants
with probability $[(1-P_k(t))/(N-1)]$. As
mentioned already, previous Monte Carlo studies
[9, 11] suggested $f \simeq 0.80$ and average
$\tau \le  10$ (see [20], reporting a slightly
different value of $f$). We first checked (for $N$
values up to 6400) that our previously reported
results for this case, the saturation value
$f_s$  of the utilization fraction $f$ and  the
convergence time $\tau$ are consistent
with our previous reports (see Figs.~\ref{fig-fasim} and~\ref{fig-tasim}).
\\
\\b) For GCA, where out of the $n_k(t-1)$ people, the person
got food in the $k$th restaurant yesterday ($t -1$),
goes back to the same $k$th restaurant today with
100$\%$ probability and the rest of $n_k -1$ agents
follow the old strategy of CA to go back to the
$k$th restaurant with  probability $P_k(t-1)$ as
defined earlier and to any one of the rest $N-1$
restaurants with probability $(1 - P_k)/(N-1)$ as in
CA. Hence, with this GCA strategy, once one of the
restaurants get a customer, that restaurant never
becomes free again (as in CA), but continues with
with the same customer or another one for the rest
of the time. This assures full utilization ($f_s = 1)$
in convergence time $\tau$ (see Fig.~\ref{fig-fanir}). This is
because, random choice of the $N$ restaurants with
equal probability suggests~\cite{chakrabarti2009kolkata} that the probability,
$P(n)$ of choosing  one restaurant by $n(\le N)$
players is given by $P(n) = (1/n!)\exp(-1)$, for
$N \to \infty$, as mentioned in the introduction.
This gives  the convergence time $\tau$ for each
of the $N$ agents to choose  distinctively
different $N$ restaurants will be equal to
$eN$, where $e$ is the Euler number (see Fig.~\ref{fig-tanir}).

\begin{figure}[!htbp]
\includegraphics[height=7cm]{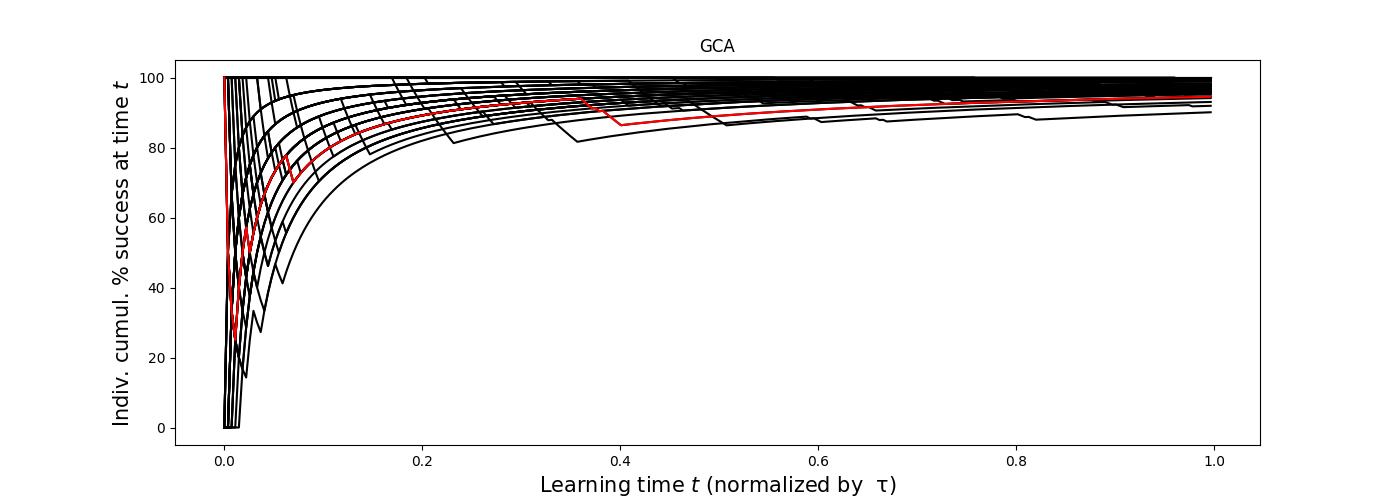}
 \caption{ 
 A typical time evolution of the cumulative
success rate of the individual agents, employing
GCA learning, starting from the first day ($t=1$)
until the convergence time ($t = \tau$) when every
one just gets food. The ``World Lines'' are for
50 agents and 50 restaurants ($N = 50$). 
As expected, with learning in GCA, this dispersion
(among the agents) in the cumulative success rate
start decreasing as the learning time $t$ $(< \tau$)
grows. For example,  if in the first 3 steps an agent
does not get food, next two times gets food, then that agent's cumulative \%
successes are 0 at times $t = 1,2,3$, then $25\%$ at time $t = 4$ and
$40\%$ at time $t = 5$). We stop these ``World Lines'' at
$t = \tau$ and find the eventual dispersion in cumulative
success to be much less than expected. We show
(in red) the ``World Line'' of a typical agent, who had initial bad luck
of successive failures eventually ending up in
about 90\% cumulative success. It is interesting
to note that these ``World Lines'' of the cumulative
success rates of the agents tend to cross
each other before the convergence time $\tau$
(mostly within the time range $0.2 \tau < t < 0.1 \tau$).  }
 \label{fig5}
\end{figure}

\begin{figure}[!htbp]
\includegraphics[width=12.5cm]{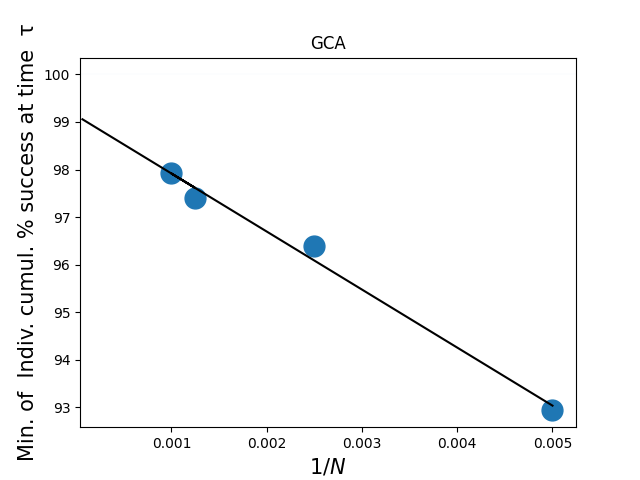}
 \caption{ 
Plot showing the variation of the agent to agent dispersion in cumulative success rate over learning time $\tau$. The minimum of 
the cumulative success decreases a bit with the system size $N$ and this dispersion limits to about 8 to 11 percent below the highest $100_{-}$
percent in the $N \to \infty$ limit. }
 \label{fig6}
\end{figure}

We further studied, for the GCA strategy,
the success rate of the customers throughout
their learning period (of duration $\tau$, the
convergence time from which onward all the
customers get food). Fig.~\ref{fig5} shows a typical
time evolution of 50 agents and 50 restaurants
($N = 50$), starting form the first day ($t = 0$)
when (for random choice restaurants) about 63$\%$
gets food and the rest about 37$\%$ fails (for
the random choice case~\cite{chakrabarti2009kolkata}  discussed in
the Introduction). With learning in GCA, this
dispersion in the cumulative success rate over
(learning) among the agents start decreasing
as can be seen from Fig.~\ref{fig5}. After convergence
time $\tau$ of course this dispersion will
gradually disappear as then every one will
be successful there. We stop at the learning
time or convergence time $\tau$ to see  how
the time average success rate for each of the
individuals (50 in number in Fig.~\ref{fig5} to give
typical picture) change with time $t$ up to
$\tau$. It is interesting to note that these
``World Lines'' of the cumulative success rates
of the agents cross each other  before the
convergence time $\tau$ (mostly within the
time range $0.2 \tau < t < 0.1 \tau$ in
Fig.~\ref{fig5})  and the eventual time average success
rate at time $ t = \tau$ does not have
much dispersion (maximum 15 from the highest
possible $100_-\%$ success rate, as in Fig.~\ref{fig5}
for  $N = 50$). In Fig.~\ref{fig6}, we show that this
agent to agent dispersion in cumulative success
rate over learning time $\tau$ also decreases
a bit with the system size $N$ and this
dispersion limits to about 9 to 13 percent below
the highest $100_-\%$ in the $N \to \infty$
limit.

 \section{Summary \& Discussion}
In the Kolkata Paise Restaurant problem,
where in general each of the $N$ agents
(or players) choose independently every
day (updating their strategy based on
past experience of failures) among the
$M$ restaurants, where he/she will be
alone or lucky enough to be picked up
randomly from the crowd who arrived at
that restaurant that day, to get the
only food plate served there (both $N$
and $M$ are macroscopically large). 
The objective of the agents is to learn themselves in
the minimum ($N$-independent) learning or convergence
time $\tau$ to have maximum success. When KPR
(with $M = N$) is used for sociological modeling (see
e.g.,~\cite{sen2014sociophysics}), the competition  with a dictated  society
(achieving full utilization $f$ = 1 instantly i.e. $\tau$ = 0),
becomes tough for the cases with individual decisions and choices by the agents themselves (achieving only $f \simeq$ 0.63 in time
$\tau$ = 0, for random choices by the agents~\cite{chakrabarti2009kolkata}). For
any reasonable competitiveness the KPR models of
``democratic'' societies with agents' individual choices need
to achieve a sufficiently high level of utilization fraction $f$,
in finite ($N$-independent) learning or convergence time
$\tau$ (see e.g.,~\cite{ghosh2010statistics} and Figs.~\ref{fig-fasim} and~\ref{fig-tasim} here).

Here we consider the most studied problem where
$M = N$. The utilization fraction $f$ of the restaurants
then also becomes equal to the success probability of
each agent. Agents here make their own  choices based
on their only available information about yesterday's
crowd size in the very restaurant visited by the
individual agent.  We show here, with single step
memory or information only of yesterday's crowd size
in the very restaurant visited by the individual,
the maximum possible utilization that can be achieved
is about eighty percent ($f \simeq 0.80$) in an optimal
time $\tau$ of order ten, even when $N$ the number of
customers or of the restaurants goes to infinity (see
Figs.~\ref{fig-fasim} and~\ref{fig-tasim}). Of course, these are numerical results
and we do not have any theoretical argument yet in its favor.
For the rest of the strategies  discussed here (see also
discussions later), while the maximum utilization
($f \rightarrow 1$) is possible, the convergence time
$\tau$ tends to infinity (see e.g., Fig.~\ref{fig-tanir}) linearly with
$N$ (or even as $\log N$).

As shown already~\cite{chakrabarti2009kolkata} (see also~\cite{chakrabarti2021development} and~\cite{harlalka2023stability}
), if each agent chooses randomly the
restaurants, then $f=1-\exp(-1) \simeq
0.63$ and $\tau = 0$. In the crowd avoiding
(CA) strategy, each agent remembers
the crowd size $n_k(t-1)$ at the $k$th
restaurant chosen and arrived at yesterday.
When those $n_k(t-1)$ agents choose the same
$k$th restaurant today with probability
$P_k(t) = 1/[n_k(t-1)]^{\alpha}$, with
$\alpha > 0$) and chooses any other of
the $N-1$ restaurants with probability
$[(1-P_k(t))/(N-1)]$, then the extentions of our previous~\cite{ghosh2010statistics,sinha2020phase}
Monte Carlo studies give 
 $f \simeq 0.80$ in average convergence
 time $\tau \le 10$  
(independent of $N$) for $\alpha$ = 1 (see Figs~\ref{fig-fasim} and~\ref{fig-tasim}). For $\alpha
\rightarrow$ 0, the Monte Carlo study in~\cite{sinha2020phase} indicated that
while the utilization fraction $f \rightarrow 1$, the
convergence time $\tau \rightarrow \infty$ as $1/\alpha$
(independent of $N$). It may be mentioned at this point that
these results are again for a single step memory of the
last visited restaurant crowd size. For the case where
the memory is for all the restaurant crowd sizes for all
the past days, one can get again full utilization ($f = 1$),
taking~\cite{dhar2011emergent,rajpal2018achieving} $\log N$ (strictly bounded by $N$) order
convergence time $\tau$. There are also some unpublished
results for some similar but a little more tricky choices
of the restaurants by the agents, but having additional
option for the choices by the restaurants , one can get
$f =$ 1 in convergence $\tau$ strictly bounded by  $N$.
However, no such choice of customer option has so far been
allowed in our models discussed here (significantly allowed
for by both the agents and the  taxis in the online hire
mobility markets~\cite{martin2017vehicle,martin2019,yang2018mean}).

Here we show that using a slightly modified
crowd avoiding algorithm, called here Greedy
Crowd Avoiding Algorithm or GCA,  where out
of the $n_k(t-1)$ agents arriving the $k$th
restaurant yesterday, the one who got
the food there yesterday, go to the $k$th
restaurant with certainty, while the other
$n_k(t-1)-1$ go to that $k$th restaurant
the probability $P_k(t-1)$  and to the other
$N-1$ restaurants with total  probability
$1-P_k(t-1)$ as in CA. The utilization fraction
$f$ then becomes unity (see Fig.~\ref{fig-fanir}), but the
convergence time $\tau$ grows linearly with $N$
($\tau = eN$; $e$ denoting the Euler number; see
Fig.~\ref{fig-tanir} and the arguments given there).

A typical time evolution of the cumulative
success rate of the individual agents, employing
GCA learning (see Fig.~\ref{fig5}) indicates that
the eventual dispersion in cumulative
success to be much less than expected. Also,
Fig.~\ref{fig6} shows that the variation of the agent
to agent dispersion in cumulative success rate
over learning time $\tau$ is limited to
about 8 to 11 percent below that of the
highest $100_{-}$ percent compared to the most
successful one  in the $N \to \infty$ limit.

This study  perhaps indicates  that using non-dictated
strategies for KPR, full utilization can never be collectively
learned or achieved in finite convergence time, when $N$,
the number of customers or of restaurants, goes to infinity,
and best compromised solution for this $N$ agent  and $N$
restaurant Kolkata Paise Restaurant problem is obtained
using the Crowd Avoiding (or CA) strategy discussed here, 
where the utilization fraction $f \simeq 0.8$ is achieved in
learning or convergence time $\tau$ of order ten even in
the large $N$ limit.

\textbf{\section*{Acknowledgement}}
We are thankful to the referees for careful reading and important suggestions. BKC is grateful to Indian National Science Academy for their Senior Scientist Research Grant.

\end{document}